\documentstyle[12pt,epsf,epsfig,wrapfig]{article}
\begin{document}

\begin{center}
{\bfseries KINEMATIC AZIMUTHAL ASYMMETRIES AND LAM-TUNG RELATION}

\vskip 5mm
O.V. Teryaev$^{1 \dag}$

\vskip 5mm
{\small
(1) {\it
BLTP. JINR
}\\
$\dag$ {\it
E-mail: teryaev@theor.jinr.ru
}}
\end{center}

\vskip 5mm
\begin{abstract}
The origin of Lam-Tung relation for the angular asymmetry
of Drell-Yan dileptons is analyzed. The asymmetry constrained by
this relation
is shown to belong  to the class of kinematic azimuthal asymmetries,
emerging due to the
deviation of the reference axis from the natural physical one.
The validity and violation of
Lam-Tung relation due to radiative and power QCD corrections
is discussed. The relation is generalized to the case when the virtual
photon is not the transverse one and the possible applications are suggested.

\end{abstract}

\vskip 8mm

The azimuthal asymmetries of dilepton pairs play a prominent role
in high energy and nuclear physics. They, from one side, may be relatively easy  studied experimentally, and
from the other side, are directly related to the virtual photon tensor
polarization and serve as a probe for the production dynamics.   
The azimuthal asymmetries of Drell-Yan pairs in hadronic collisions are the subject of theoretical and experimental studies for 
many years ( see \cite{BBN} and ref.therein). There are various competing dynamical mechanisms which may contribute to asymmetry.
The recent proposal of studies of very high mass dileptons by CMS detector at LHC opens the exciting possibility 
of investigating the asymmetries in the entirely new kinematical domain and disentangle various dynamical mechanisms by their 
energy and dilepton mass dependence.     

The angular distribution of dileptons has the following form  
\begin{eqnarray}
\label{angledist}
        d\sigma \propto
        1+\lambda \cos^2\theta +\mu \sin 2\theta\cos\phi
        +{\nu \over 2}\sin^2\theta\cos 2\phi
\end{eqnarray}
where $\theta$ and $\phi$ are polar and azimuthal angles of one of leptons in their c.m. frame, while
$\lambda,\ \mu,\ \nu$ are angle independent coefficients encoding the 
information on tensor polarization of virtual photon.  

The special role in the analysis of azimuthal asymmetries is played by Lam-Tung (LT) relation \cite{LT}.   
\begin{eqnarray}
\label{LT}
        \lambda + 2 \nu =1
\label{lt}
\end{eqnarray}

This relation, surviving  the account for leading order QCD corrections, plays the prominent role in the analysis of 
dynamical mechanisms. At the same time, its physical origin remains unclear, although the field theoretical derivation 
is on the solid ground.

In the present paper it is shown, that LT relations describes the azimuthal asymmetry belonging to the class of {\it kinematical}
ones. Namely, the azimuthal asymmetry is absent at all in the natural reference frame, defined by the collision kinematics, and emerges 
solely as a kinematical effect when passing to the actual reference frame.

We start with the following expression with respect to some (variating) 
axis $m$
\begin{eqnarray}
\label{kin}
        d\sigma \propto
        1+\lambda_0 (\vec n \vec m)^2=1+\lambda_0 \cos^2\theta_{nm}^2
\end{eqnarray}
where 
\begin{eqnarray}
\label{kin1}
        \cos \theta_{nm}= \cos \theta \cos \theta_0 + \sin \theta \sin \theta_0 \cos \phi..
\end{eqnarray}
Note that $ \theta_0$ and $ \theta$ are the angles formed by the 
axis $m$ and lepton direction $n$ with coordinate axis $z$, while 
$. \phi$ is their relative azimuthal angle.

As a result one gets the azimuthal dependence with respect to this 
coordinate axis, so that

 \begin{eqnarray}       
        \lambda = \lambda_0 \frac {2-3 \sin^2 \theta_0}{2+ \lambda_0 \sin^2\theta_0}, \nonumber \\
        \nu= \lambda_0 \frac {2 sin^2 \theta_0}{2+ \lambda_0 \sin^2\theta_0}.
        \end{eqnarray}

One may exclude $\theta_0$ and get 

\begin{eqnarray}       
        \lambda_0 = \frac {\lambda+\frac{3}{2}\nu}{1- \frac{1}{2} \nu}.
        \end{eqnarray}

As a result, we found the new relation which 
is reduced to the standard LT relation in the case of transverse 
virtual photon  when $  \lambda_0 =1  $. We also see that 
LT relation accomodates two very different inputs,
namely the kinematical nature of the asymmetry and the transverse polarization
of virtual photon. This is probably the main origin of the difficulties 
in finding its physical explanation. 

For practical application of kinematical generation of asymmetry we 
need to depart from the simplest version of the model with the fixed 
direction of physical axis and to be able to sum over its various directions.
For this purpose the theorem for averages is very useful:

   \begin{eqnarray}       
        \frac {\int d \vec x \Delta \sigma (\vec x)}{ \int d \vec x \sigma (\vec x)}=\frac {\Delta \sigma (\vec x_0)}{\sigma (\vec x_0)}
        \end{eqnarray}
where $\sigma(\vec x)$ is positive and 
$\vec x_0$ is some point in the integration region. If integration is N-dimensional, these points typically form
N-1-dimensional subspace ("orbit"),
This theorem is very well suited for the studies of parton-like 
expressions for spin asymmetries and, in particular, provides the 
natural explanation  \cite{OT05} of the suppression  \cite{Ans}
of Collins function contribution to inclusive Single Spin Asymmetries 
in the hadronic processes with respect to that of the Sivers function. 

To apply this theorem in the actual case one may
introduce 
\begin{eqnarray}
\label{av}
        d\sigma = \sigma \lambda_0 (\vec n \vec m)^2.
\end{eqnarray}
One should get after the average over direction of $\vec m$
\begin{eqnarray}
\label{avf}
        d\sigma \propto 1+\lambda_0 (\vec n \vec m_0)^2.
\end{eqnarray}

Note that $\vec m_0$, generally speaking, depend on the vector $\vec n$. 
This means that one should have a corresponding set of orbits, 
which are one dimensional curves  
covering the unit sphere. 
However, they should intersect the scattering plane in the same point,
in order to avoid the appearance of the terms in the angular distribution which are not compatible with 
the general structure (\ref{angledist}).

This  justifies the applicability of kinematical 
$m-$model for the inclusive Drell-Yan processes.  
 
Let us now compare the cases of validity and violation of  LT relation
discussed in  \cite{BBN}.
In the  LO QCD the azimuthal asymmetry with respect to the physical partonic
axis is absent, leading to validity of LT relation. This is not  true
at NLO QCD, when the partonic scattering plane enables the dynamical 
azimuthal asymmetry. At the same time, the substantial part of the whole 
NLO result is due to the emission of collinear gluons where scattering 
plane is unessential, and this may explain the smallness of NLO
violation of LT relation.

The scattering plane and resulting dynamical azimuthal asymmetry,
violating the LT relation, is also present in the higher
twist contribution  \cite{BBM} to DY pairs production by pions scattered  
on nucleons.

Finally, note that the kinematical azimuthal asymmetries 
may appear for the dileptons produced in quark-gluon plasma 
and hadronic media, Their polar angular asymmetry may be different 
and  
is quite sensitive   \cite{brat} so the production mechanism.
So, the new relation 
obtained in this paper may find another natural application.


\begin{thebibliography}{99}
\bibitem{BBN}
D.~Boer, A.~Brandenburg, O.~Nachtmann and A.~Utermann,
  Eur.\ Phys.\ J.\ C {\bf 40}, 55 (2005)
  [arXiv:hep-ph/0411068].
\bibitem{LT}
C.~S.~Lam and W.~K.~Tung,
  Phys.\ Rev.\ D {\bf 21}, 2712 (1980).


\bibitem{OT05} O.V. Teryaev, talk at RHIC Spin Collaboration Meeting, Torino,
October 8-9, 2004.
 
\bibitem{Ans}
M.~Anselmino, M.~Boglione, U.~D'Alesio, E.~Leader and F.~Murgia,
  Phys.\ Rev.\ D {\bf 71}, 014002 (2005)


\bibitem{BBM}
 A.~Brandenburg, S.~J.~Brodsky, V.~V.~Khoze and D.~Mueller,
  Phys.\ Rev.\ Lett.\  {\bf 73}, 939 (1994)

\bibitem{brat}
E.~L.~Bratkovskaya, O.~V.~Teryaev and V.~D.~Toneev,
  Phys.\ Lett.\ B {\bf 348}, 283 (1995).




\end{thebibliography}
\end{document}